# Turning Up the Heat: Assessing 2-m Temperature Forecast Errors in AI Weather Prediction Models During Heat Waves

Kelsey E. Ennis, Elizabeth A. Barnes, Marybeth C. Arcodia, Martin A. Fernandez, Eric D. Maloney

*Department of Atmospheric Science, Colorado State University, Fort Collins, CO*

*Corresponding author*: Kelsey E. Ennis, kelsey.ennis@colostate.edu

ABSTRACT

Extreme heat is the deadliest weather-related hazard in the United States. Furthermore, it is increasing in intensity, frequency, and duration, making skillful forecasts vital to protecting life and property. Traditional numerical weather prediction (NWP) models struggle with extreme heat for medium-range and subseasonal-to-seasonal (S2S) timescales. Meanwhile, artificial intelligence-based weather prediction (AIWP) models are progressing rapidly. However, it is largely unknown how well AIWP models forecast extremes, especially for medium-range and S2S timescales. This study investigates 2-m temperature forecasts for 60 heat waves across the four boreal seasons and over four CONUS regions at lead times up to 20 days, using two AIWP models (Google GraphCast and Pangu-Weather) and one traditional NWP model (NOAA United Forecast System Global Ensemble Forecast System (UFS GEFS)). First, case study analyses show that both AIWP models and the UFS GEFS exhibit consistent cold biases on regional scales in the 5–10 days of lead time before heat wave onset. GraphCast is the more skillful AIWP model, outperforming UFS GEFS and Pangu-Weather in most locations. Next, the two AIWP models are isolated and analyzed across all heat waves and seasons, with events split among the model's testing (2018–2023) and training (1979–2017) periods. There are cold biases before and during the heat waves in both models and all seasons, except Pangu-Weather in winter, which exhibits a mean warm bias before heat wave onset. Overall, results offer encouragement that AIWP models may be useful for medium-range and S2S predictability of extreme heat.

SIGNIFICANCE STATEMENT

Extreme heat is a deadly hazard that traditional forecast models struggle to predict, particularly for longer lead times. While artificial intelligence-based weather prediction (AIWP) models are rapidly gaining in popularity, little is known about their ability to forecast heat waves and other extreme events. This study evaluates the temperature forecast skill of two AIWP models and one traditional forecast model (NOAA United Forecast System Global Ensemble Forecast System (UFS GEFS)) across 60 US heat waves. While all models exhibit consistent cold biases for heat waves, the Google GraphCast AIWP model offers the most promise, at times outperforming UFS GEFS. Results emphasize that AIWP models may provide substantial advances in the medium- and long-range forecasts of extreme heat events.

# 1. Introduction

Extreme heat is one of the deadliest weather-related hazards across the globe (e.g., Limaye et al. 2018; Rennie et al. 2021; Ballester et al. 2023), resulting in as many as 5,000 excess deaths per year in the U.S. alone (Weinberger et al. 2021). Excess mortality can be particularly amplified in regions with less modern infrastructure, adaptive capacity, and/or cooling access (e.g., Wu et al. 2022). For example, during a series of heat waves in Europe in 2022, Ballester et al. (2023) estimates that there were more than 60,000 excess deaths due to the cascading impacts of extreme heat. According to the IPCC Sixth Assessment (AR6) and the Fifth US National Climate Assessment (NCA), extreme heat causes adverse impacts to socioeconomic and environmental systems, including but not limited to health, ecosystems, agriculture, energy, as well as national and regional economies (IPCC 2022; USGCRP 2023). Furthermore, the Fifth NCA emphasizes that extreme heat in the US can be associated with drought (i.e., reduced water supply and agricultural yields), increased wildfire activity, and stress on the energy grid (USGCRP 2023). Power outages and blackouts during extreme heat events can also greatly increase mortality and morbidity (e.g., Stone Jr. et al. 2023). Moreover, energy demand during extreme heat events is increasing with time and the changing climate (e.g., Rastogi et al. 2021), stressing the electrical grid and in turn making power outages and blackouts more likely.

Extreme heat intensity, frequency, and duration are increasing and projected to continue to do so, both across the U.S. and globally (e.g., Perkins 2015; Mora et al. 2017; Keellings and Moradkhani 2020; Clarke et al. 2022; Barriopedro et al. 2023; Domeisen et al. 2023). To this end, Khatana et al. (2024) projects that deaths related to heat waves and extremely hot temperatures will increase substantially by 2050, particularly impacting older individuals and those with less access to cooling.

Although there is no universal definition for heat waves, they are generally characterized as periods of highly anomalous surface warmth that lasts for at least three days and can occur with or without elevated humidity levels (e.g., Perkins 2015; Barriopedro et al. 2023). Heat waves can be caused by a variety of physical mechanisms, particularly atmospheric blocking, land-atmosphere processes (e.g., anomalously low soil moisture), ocean-atmosphere coupling (e.g., anomalously warm sea surface temperatures), and topography that can set the location of stationary waves and therefore impact the shape of the jet stream (Jiménez-Esteve and Domeisen 2022; Barriopedro et al. 2023; Domeisen et al. 2023).

Traditional numerical weather prediction (NWP) model forecast skill is constantly improving, with useful forecasts (defined as anomaly correlation coefficient > 0.6; Bauer et al. 2015) being regularly produced 10–14 days into the future (Alley et al. 2019; Cahill et al. 2024). However, the predictability of heat waves and other extreme weather events is still a challenge for traditional NWP models, especially for medium-range and subseasonal-to-seasonal (S2S) timescales (e.g., Vitart and Robertson 2018; Lin et al. 2022; Barriopedro et al. 2023; Xie et al. 2024). Improving predictability of extremes on S2S timescales is crucial to establish early warning systems and increase societal readiness (Vitart and Roberston 2018), including to sectors such as health, agriculture, energy, water resources, and emergency management (Klemm and McPherson 2017; White et al. 2017; White et al. 2022). S2S forecasts have improved over time and can predict the evolution of large-scale and long-duration weather events (e.g., Vitart and Robertson 2018). However, for extreme heat events, results from traditional NWP efforts on S2S timescales are decidedly mixed. For example, Ford et al. (2018) found that a NOAA coupled model failed to capture the full duration of US heat waves. In addition, the model inaccurately represents land-atmosphere feedbacks in certain regions, degrading forecast skill, which is similarly found by Seo et al. (2024) over the Western US. Using the extended ensemble forecast system from the European Centre for Medium-range Weather Forecasts (ECMWF), Lavaysse et al. (2019) showed that cold spells are more skillfully predicted than heat waves on S2S timescales, although both phenomena become much less predictable at lead times of greater than two weeks. Barriopedro et al. (2023) stated that misrepresentation of coupled land-atmosphere, diabatic, and/or convective processes, as well as model biases with respect to large-scale circulation patterns, can adversely impact S2S forecast skill of extreme heat. Overall, there is much room for improvement to S2S forecasts of extreme heat, and emerging hope that newer artificial intelligence-based weather prediction (AIWP) models can offer substantial advances in predictive skill (Pasche et al. 2025).

AIWP models are increasingly being used in atmospheric science research and operational weather forecasting, as they require a fraction of the required computational power compared to traditional NWP models (Bouallègue et al. 2024; Radford et al. 2025). AIWP models are proving to be skillful forecast tools, especially on larger spatial scales and in some cases longer lead times (e.g., Bouallègue et al. 2024; Waqas et al. 2024; Xie et al. 2024). For example, Xie et al. (2024) diagnose the evolution of heat waves in China multiple weeks in advance, by training a convolutional neural network (CNN) model to utilize extreme heat precursors. Along those lines, Lopez-Gomez et al. (2023) use a set of neural weather models to forecast global surface temperature anomalies, finding significant skill improvement compared to traditional S2S NWP models up to 28 days in advance. Meanwhile, Li et al. (2023) train a graph neural network model on surface station data across the CONUS and use it to make skillful forecasts of regional heatwaves. Overall, AIWP models can produce forecasts that compete with traditional operational NWP models (Hakim and Masanam 2024), offering promise for their future in predicting key weather elements such as temperature and precipitation (e.g., Waqas et al. 2024).

Despite their potential, AIWP models have numerous limitations, including their inability to represent fundamental dynamic and thermodynamic processes (Selz and Craig 2023; Bonavita 2024), struggles to accurately simulate mesoscale weather features (Bonavita 2024), producing overly smooth forecasts and increasing biases with time (Bouallègue et al. 2024), and failure to reproduce the butterfly effect associated with atmospheric chaos (i.e., the model forecasts are not sensitive enough to initial conditions; Selz and Craig 2023). AIWP models also require a more intense verification by the scientific community, particularly for forecasts of extreme events (Pasche et al. 2025; Radford et al. 2025; Ullrich et al. 2025). As an example, Pasche et al. (2025) examine three recent extreme events (two heat waves) using four AIWP models. Specifically for the 2021 Pacific Northwest and 2023 South Asian heat waves, they find similar accuracy between the AIWP models and a leading traditional NWP model. However, the AIWP models lack small scale details and the appropriate variables to conclusively diagnose the extreme heat events.

This study aims to systematically evaluate the ability of two AIWP models (Google GraphCast and Pangu-Weather, hereafter GraphCast and Pangu) to predict heat waves in all four boreal seasons across the CONUS and compare their skill to a traditional S2S NWP model, the NOAA United Forecast System Global Ensemble Forecast System (UFS GEFS). We focus our evaluation on medium-range and subseasonal prediction timescales (out to 20 days). First, surface temperature forecasts from the two AIWP models and UFS GEFS are evaluated and compared for two heat wave case studies across disparate CONUS regions, allowing us to investigate model skill and sensitivity at regional levels. Subsequently, the regional and seasonal temperature forecast skill of the two AIWP models is explored across a larger group of heat waves, during and outside of the model training period, allowing for a robust evaluation of model performance during extreme heat events. Our primary objective is to assess—specifically for heat waves across the CONUS—if, where, and when the two AIWP models show promise or offer advantages over a traditional NWP system.

## 2. Data and methods

*a. Traditional NWP model and verification dataset*

For the traditional NWP model, we use the NOAA GEFSv12, an uncoupled version of the UFS (hereafter UFS GEFS) produced in September 2020 (Guan et al. 2022). Variables from the control run are available on a 6-h basis at 0000, 0600, 1200, and 1800 UTC. The reforecasts are run out to lead times of 1–35 days (inclusive), leading to a total of 1042 samples initialized every seven days from 5 January 2000 to 18 December 2019. To create daily mean temperatures, we average these 6-h instantaneous forecasts for each day over the entire dataset. NOAA provides UFS GEFS data for lead times of Days 11–35 which are produced at half the grid spacing (0.5° x 0.5°) of the data for lead times 1–10 (0.25° x 0.25°). Therefore, prior to merging the datasets for calculations, we regrid output fields for lead times 11–35 using the python module xESMF's bilinear interpolation (Zhuang et al. 2024), such that all lead times have a 0.25° x 0.25° grid spacing. This ensures that all forecast lead times share the same spatial coordinates, allowing for comparisons and calculations without spatial mismatch.

To calculate errors in model forecasts and identify large scale heat waves (section 2c), we use the ECMWF ERA5 reanalysis (Hersbach et al. 2020) as our ground truth. ERA5 is produced on a reduced Gaussian grid, with a quasi-uniform grid spacing of approximately 0.25° x 0.25° and has hourly data from 1940–present. To obtain daily 2-m temperature fields, we aggregate the 6-h 2-m temperature output to 24-h resolution. While the full ERA5 dataset goes back to 1940, the two AIWP models (section 2b) are trained on ERA5 data from 1979–2017.

*b. AIWP models*

GraphCast, an AIWP deep-learning model, is based on a graph neural network (GNN) architecture following an "encode-process-decode" configuration (Lam et al. 2023). GNN-based learned simulators are successful at capturing complex fluid dynamics and other governing partial differential equations, making them well equipped for weather modeling (e.g., Lam et al.

2023). GraphCast is autoregressive, meaning it can be "rolled out" by feeding its own predictions back in as input, to generate longer trajectories of weather states (Lam et al. 2023). The model includes four surface variables (2-m temperature, mean sea-level pressure, 10-m u-component of the wind, 10-m v-component of the wind) and five additional variables (temperature, geopotential height, specific humidity, u-component of the wind, v-component of the wind) on 37 pressure levels, outputting variables on a 6-h basis.

Pangu is also an AIWP deep learning model and incorporates a 3D Earth-specific transformer (3DEST) deep network architecture, which allows the model to represent spatial dependencies through self-attention mechanisms (Bi et al. 2023). The model outputs five variables (temperature, geopotential height, specific humidity, u-component of the wind, v-component of the wind) on 13 pressure levels; at the surface, there are four output variables (2-m temperature, mean sea-level pressure, 10-m u-component of the wind, 10-m v-component of the wind). Pangu has four trained networks, each optimized for different forecast intervals: 1-h, 3-h, 6-h, and 24-h. For medium-range forecasting, the Pangu model developers introduced hierarchical temporal aggregation. As a part of this approach, developers suggest using the 24-h forecast model when running forecasts out to longer lead times to minimize the number of iterations required (Bi et al. 2023). However, we use the Pangu 6-h model to remain consistent with UFS GEFS and GraphCast output. This ensures that we can compute consistent daily averages for each lead time across all the models for accurate comparison. We discuss this choice in more detail in section 4.

*c. Heat wave identification*

For a thorough investigation of the 2-m temperature prediction skill of the UFS GEFS and AIWP models, we investigate a total of 60 heat waves (15 per season) during 2000–2023. These heat waves span the AIWP model training period (1979–2017) and extend past it into the model testing years (2018–2023). We select 60 heat waves to ensure that we have a large enough sample size per boreal season to be able to divide events into two groups: heat waves in the model testing years and heat waves in the training years. There are seven total heat waves in each season that are in Pangu and GraphCast's testing years and eight that occur during the training years. Seven events in each season provides a focused dataset for evaluating the predictive skill of the models in the testing years. The heat waves in the six testing years offer insight into their prediction skill post-training, although we acknowledge that seven heat waves per season is still a limited sample size.

To produce our heat wave event database, we identify heat waves during boreal summer (JJA), autumn (SON), winter (DJF), and spring (MAM) for 2000–2023. Heat waves are identified using ERA5 daily mean 2-m temperature data. There is no universal definition of heat waves (Perkins 2015); however, many studies stipulate a percentile-based threshold approach with a minimum duration of three days (e.g., Perkins 2015; Cloutier-Bisbee et al. 2019), which we apply here. We choose the 95th percentile of 2-m temperatures in each season as our threshold. A heat wave event at each grid point is identified when the 95th percentile is exceeded for three or more days. Using the 98th and 99th percentiles result in some similar heat waves but provides us with too few events per season.

To identify large-scale heat waves, we use the NCA regions (USGCRP 2023, Fig. 1). We focus on heat waves within four disparate NCA regions: the Northwest, Southeast, Midwest, and Northeast (Fig. 1). We require each heat wave to have at least 100 contiguous grid points (a 2.5° x 2.5° box if a square) exceeding the 95th percentile of seasonal 2-m temperatures on a given day. This ensures that each heat wave affects a significant portion of the CONUS. As an example, Fig. 2 shows results for the record-shattering June 2021 Pacific Northwest heat wave (Schumacher et al. 2022; White et al. 2023). In Fig. 2a, only a few grid points exceed the 95th percentile threshold on 27 June 2021. One day after (28 June 2021, Fig. 2b), there is a large area that exceeds our percentile and spatial thresholds over much of the Northwest NCA region; as such, we define the start day of the heat wave as 28 June 2021, consistent with other work (e.g., Schumacher et al. 2022). We allow a heat wave event to exist in multiple NCA regions, simultaneously and/or over the course of the event. For example, a large heat wave from 30 June to 2 July 2012 affected parts of the Midwest and Southeast regions (Fig. S1).

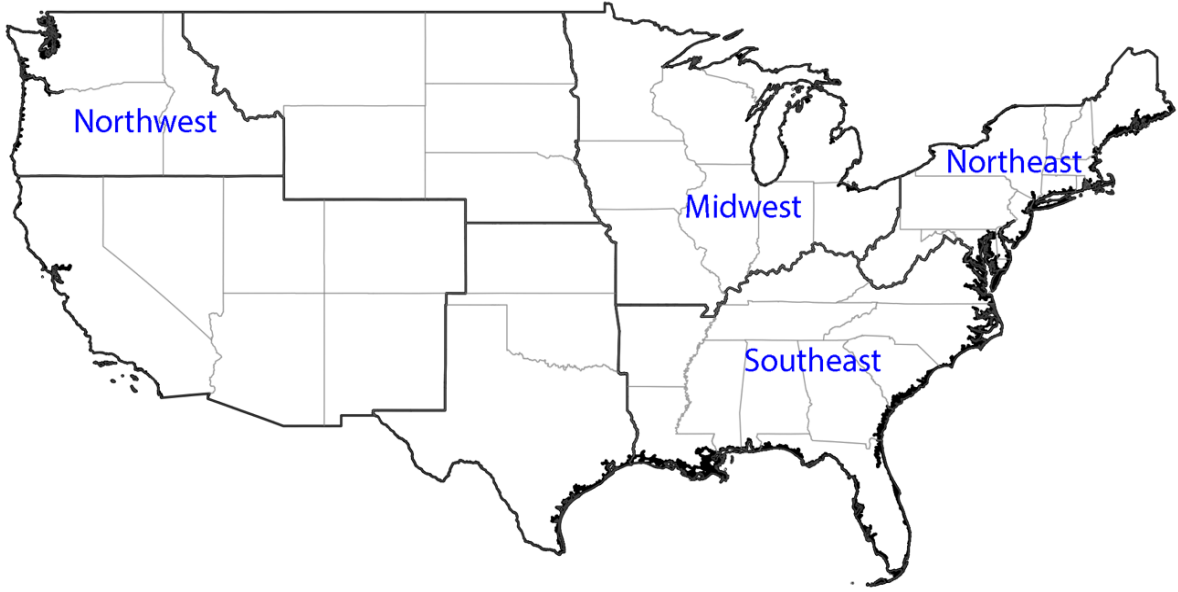

Fig. 1. Map of the seven NCA regions (black bold borders) over the CONUS, with the four regions used in this study (Northwest, Midwest, Northeast, Southeast) labeled in blue. Adapted from USGCRP (2023).

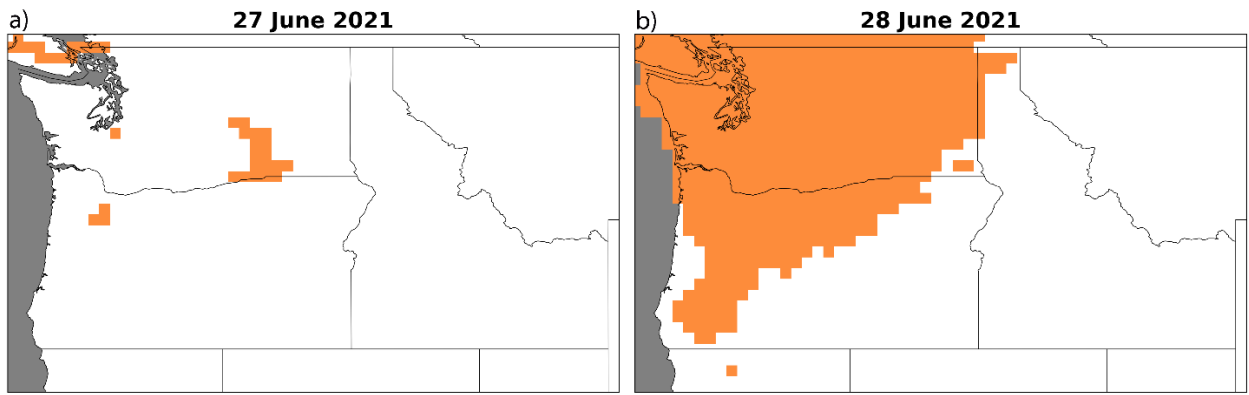

Fig. 2. The June 2021 record-shattering heat wave in the Northwest NCA region (Washington, Oregon, Idaho), where the orange shading indicates areas where daily mean 2-m temperatures exceeded the 95th percentile on (a) 27 June and (b) 28 June.

Finally, we emphasize that anomalous warm events outside of boreal summer have no standard name. Some studies (e.g., Schwarz et al. 2020) refer to them as "extreme heat", while other sources such as ECMWF term them "warm spells" (ECMWF 2020). Regardless of specific terminology, our definition of heat wave events (exceeding the 95th percentile of 2-m temperature with a minimum duration of three days over at least 100 adjacent grid points) remains consistent across all four boreal seasons and the four NCA regions. Therefore, for the remainder of this paper we use the term "heat wave" regardless of region and season.

*d. Model comparison*

In section 3a, we investigate how 2-m temperature forecast error for both AIWP models, GraphCast and Pangu, compares to that of the UFS GEFS during heat waves. We examine two case studies to demonstrate the performance of each model; specifically, we select the August 2011 Southeast boreal summer and 2019 Pacific Northwest boreal autumn heat waves. These heat waves are chosen for our case studies because 1) they occur in two different seasons, and 2) the 2011 Southeast heat wave falls within the AIWP model training years, while the 2019 heat wave occurs during the testing years. Furthermore, because it is only initialized every seven days, UFS GEFS has specific forecast dates from which to choose. Thus, we must be intentional when choosing UFS GEFS forecast initialization dates to ensure that the reforecast period covers the majority of days during each heat wave. For the August 2011 heat wave, we use the UFS GEFS run initialized on 27 July 2011, seven days prior to the onset of the heat wave event. For the September 2019 event, we choose the run initialized on 28 August 2019, four days prior to the onset of the heat wave. We then generate 20-day GraphCast and Pangu hindcasts, initializing them on the same date as the UFS GEFS reforecasts. ERA5 is used as the relevant "ground truth" baseline throughout each event's forecast period to compare performance.

In section 3b, we isolate the two AIWP models and systematically assess their subseasonal 2-m temperature prediction skill. Unlike UFS GEFS, GraphCast and Pangu do not have specific weekly initialization dates, enabling us to initialize hindcasts at any date. We initialize these hindcasts such that lead time 10 of the model forecast is the first day of the heat wave. Hindcasts are run out to 20 days.

## 3. Results

*a. Case study evaluation*

For the August 2011 Southeast heat wave, the UFS GEFS maintains low errors throughout the entire forecast, despite a slight warm bias in some locations (Fig. 3). UFS GEFS outperforms Pangu at all lead times but exhibits only slightly smaller errors than GraphCast through day 12 (Fig. 3). At a lead time of 7 days (3 August 2011; the first day of the heat wave), UFS GEFS (Fig. 4b) matches well with ERA5 (Fig. 4a), while both GraphCast (Fig. 4c) and Pangu (Fig. 4d) exhibit a cool bias throughout the Southeast. However, the Pangu cool bias is substantially larger than that of GraphCast. We choose lead time 7 for analysis (3 August 2011) because it is the first day of the heat wave and the hottest day in terms of the Southeast regional average temperature (Fig. 3a). To that end, of the three models, only UFS GEFS captures the widespread intensity of the heat wave across the region (Fig. 4). However, an encouraging aspect of the forecasts is that all three models capture the relatively cooler temperatures in the mountains of western North Carolina and eastern Tennessee, albeit with slightly varying skill (Fig. 4).

The hindcasts in Fig. 3 illustrate forecast performance evolution leading up to and through the duration of the August 2011 Southeast heat wave. UFS GEFS is the best of the three forecast models prior to and during the early part of the heat wave, with temperature errors near zero (Fig. 3b). GraphCast is the better of the two AIWP models and exhibits smaller errors than UFS GEFS toward the end of the heat wave (Fig. 3). In contrast, Pangu exhibits the largest errors and inconsistency throughout much of the heat wave, with a persistent cold bias.

The September 2019 Northwest heat wave (Figs. 3c,d) features larger temperature fluctuations and errors across all models compared to the 2011 Southeast event. Large warm biases in UFS GEFS boreal summer 2-m air temperature forecasts are demonstrated over the Northwest by Seo et al. (2024), and agree with our results, especially prior to and during the first half of the heat wave (Fig. 3d). In contrast to the UFS GEFS warm bias, GraphCast exhibits a cool bias leading up to the 2019 heat wave, lower error during much of the event, and a prominent warm bias near and just after the end of the heat wave (Fig. 3d). Pangu errors are larger than GraphCast but competitive with UFS GEFS; Pangu is particularly inconsistent prior to and during the first half of the heat wave (Fig. 3d). Overall, while GraphCast exhibits smaller errors (by 2–3°C) than UFS GEFS during the heat wave period, both GraphCast and Pangu have warm biases near the end of and just after the heat wave that exceed the warm bias in UFS GEFS by 2–4°C (Fig. 3d).

We next examine temperature errors for the August 2011 heat wave at three distinct locations within the Southeast NCA region (Fig. 1): Asheville, NC; Little Rock, AR; and New Orleans, LA. These locations are chosen based on their differing geographical characteristics (i.e., coastal proximity and elevation). All three models exhibit a cold bias for nearly the entire period at all three cities (Figs. 5b,d,f), except for Pangu near the end of the heat wave at Little Rock (Fig. 5f).

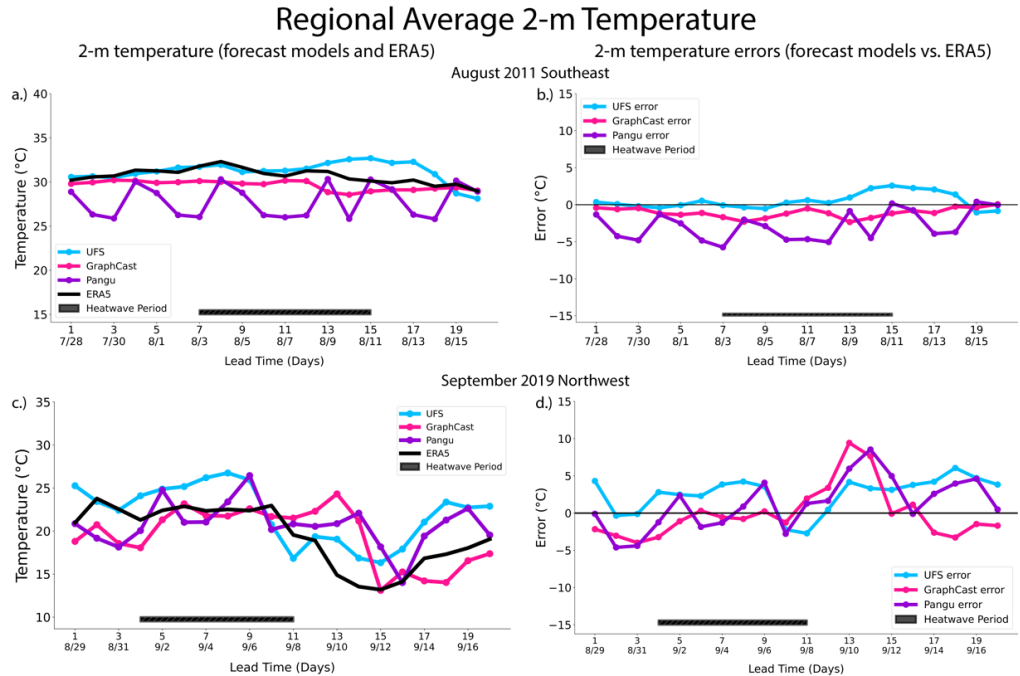

Fig. 3: Regional average (left) 2-m temperature (°C) and (right) 2-m temperature errors (°C) in UFS GEFS (blue line), GraphCast (magenta line), and Pangu (purple line) for the (a,b) August 2011 Southeast heat wave and (c,d) September 2019 Northwest heat wave. Temperature errors are calculated using ERA5 (black line on left-hand panels) as truth and are shown as a function of time. The heat wave period is illustrated by the black line at the bottom of each panel.

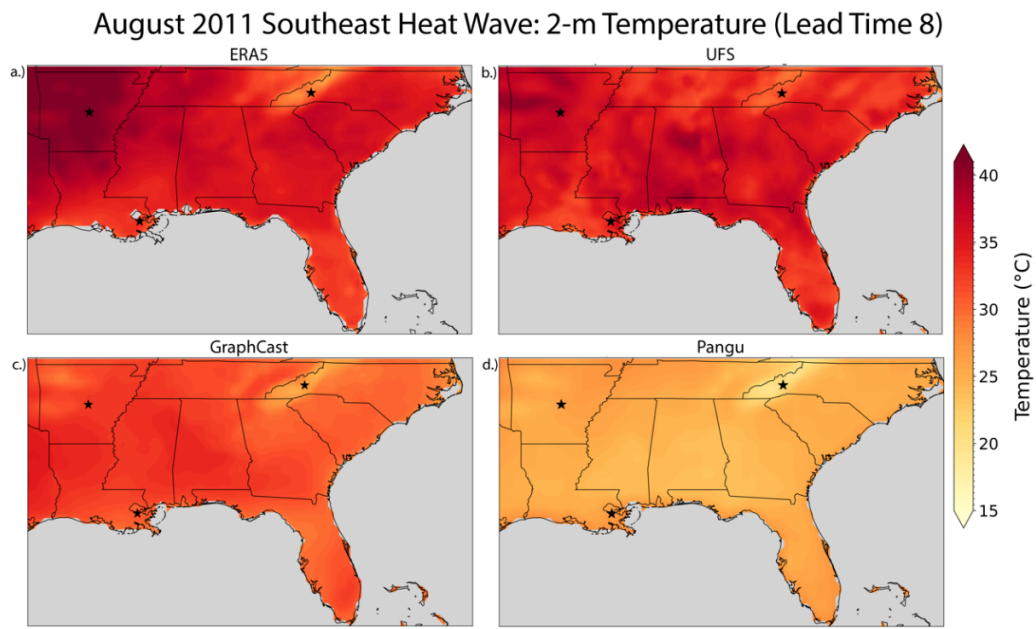

Fig. 4: For the August 2011 Southeast heat wave valid at lead time 7 (3 August 2011; the first day of the heat wave): 2-m temperature (°C, shaded) for (a) ERA5 verification; and forecasts from (b) UFS GEFS; (c) GraphCast; and (d) Pangu. The black stars (Asheville, NC; New Orleans, LA; Little Rock, AR) denote the three cities selected for analysis in Fig. 5.

## Gridpoint 2-m Temperature: August 2011 Southeast Heat Wave

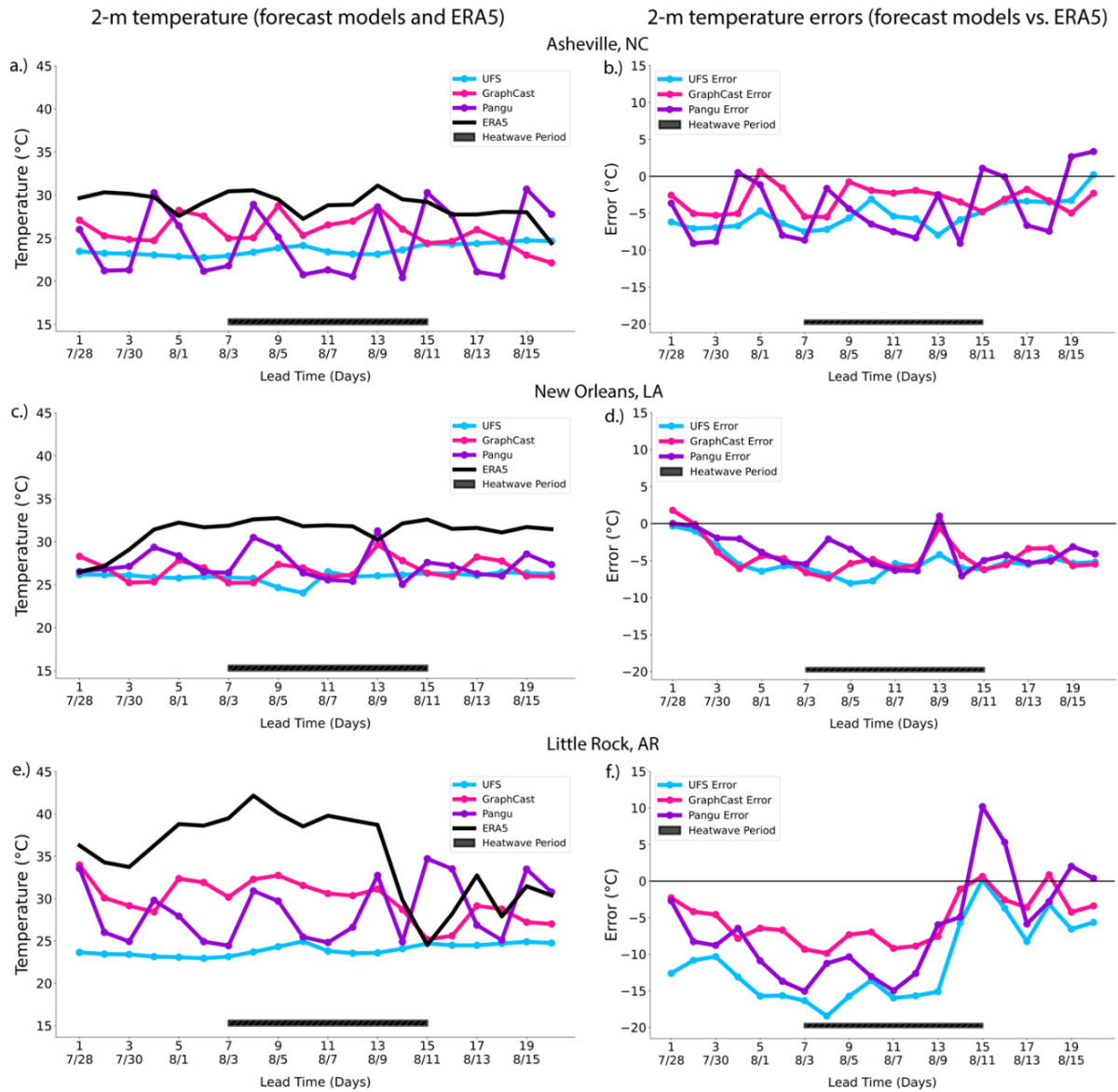

Fig 5: Grid point values of (left) 2-m temperature (°C) and (right) 2-m temperature errors (°C) in UFS GEFS (blue line), GraphCast (magenta line), and Pangu (purple line) for three locations during the August 2011 Southeast heat wave: (a,b) Asheville, NC; (c,d) New Orleans, LA; (e,f) Little Rock, AR. Temperature errors are calculated using ERA5 (black line on left-hand panels) as truth, and are shown as a function of time. The heat wave period is illustrated by the black line at the bottom of each panel.

GraphCast consistently outperforms both UFS GEFS and Pangu at the highest elevation location (Asheville; Fig. 5b). Meanwhile, the largest errors in all three models occur at Little Rock (Fig. 5f), especially before and during the first half of the heat wave. While Asheville is at a much higher elevation (650 m) than Little Rock (102 m), both western North Carolina and Arkansas have large topography gradients. Therefore, it is plausible that the coarse grid spacing of UFS GEFS limits its ability to accurately predict 2-m temperature extremes in such regions.

At New Orleans (Fig. 5d), UFS GEFS has a slightly larger cold bias compared to both AIWP models. This is notable because New Orleans is located in a flat area near or below sea level, and adjacent to the Gulf. Therefore, its temperature is regulated by its proximity to the Gulf, is not impacted by orographic effects, and generally has lower variability than locations farther inland.

We next investigate temperature errors in Seattle, WA; Bend, OR; and Boise, ID for the September 2019 Northwest heat wave (Fig. 6). As for the August 2011 case, we examine grid points with varying geography. UFS GEFS exhibits a consistent cool bias in all three locations (Figs. 6b,d,f), mirroring results for the August 2011 Southeast event, and is perhaps also related to the coarse resolution limiting accuracy in regions of complex terrain and topography gradients. GraphCast maintains the lowest errors for all three locations before and during the heat wave period. However, the model exhibits a consistent warm bias and higher error after the heat wave ends (Figs. 6b,d,f). Pangu performs well compared to UFS GEFS in all three locations but exhibits much larger error variability in Boise (Fig. 6f) than in Seattle (Fig. 6b) and Bend (Fig. 6d). All three models perform the worst in Boise (Fig. 6f), which of the three cities is the location with the highest elevation and most complex terrain. While both AIWP models outperform UFS GEFS at the three specific Northwest locations, this is not the case for the regional averages (Fig. 3).

*b. AIWP model performance*

We next examine the regional average performance of GraphCast and Pangu during four heat waves that affected four different NCA regions (Northwest, Midwest, Northeast, Southeast). Figures 7a and 7b show results for the June 2021 record-shattering Northwest heat wave (e.g., Schumacher et al. 2022; White et al. 2023). Pangu exhibits a persistent cold bias with large variations (Fig. 7b), matching the regional average results for the September 2019 event (Fig. 4). The cold bias is particularly pronounced during the early and middle portions of the heat wave (Fig. 7b). GraphCast errors are minimal before heat wave onset; however, there is a consistent cold bias during the heat wave, albeit a smaller one than Pangu (Fig. 7b). The average GraphCast absolute error over the 20-day forecast for this event (3.2 °C) is the largest of the four events in Fig. 7.

For the September 2023 Midwest heat wave (Figs. 7c,d), GraphCast consistently outperforms Pangu. GraphCast error is nearly zero until heat wave onset, followed by a slight cold bias during the heat wave and a small warm bias after the heat wave (Fig. 7d). The average GraphCast absolute error (1.4 °C) is the lowest of the four events in Fig. 7. Pangu exhibits the same general error pattern as GraphCast but is more inconsistent and has a much larger cold bias during the heat wave (Fig. 7d).

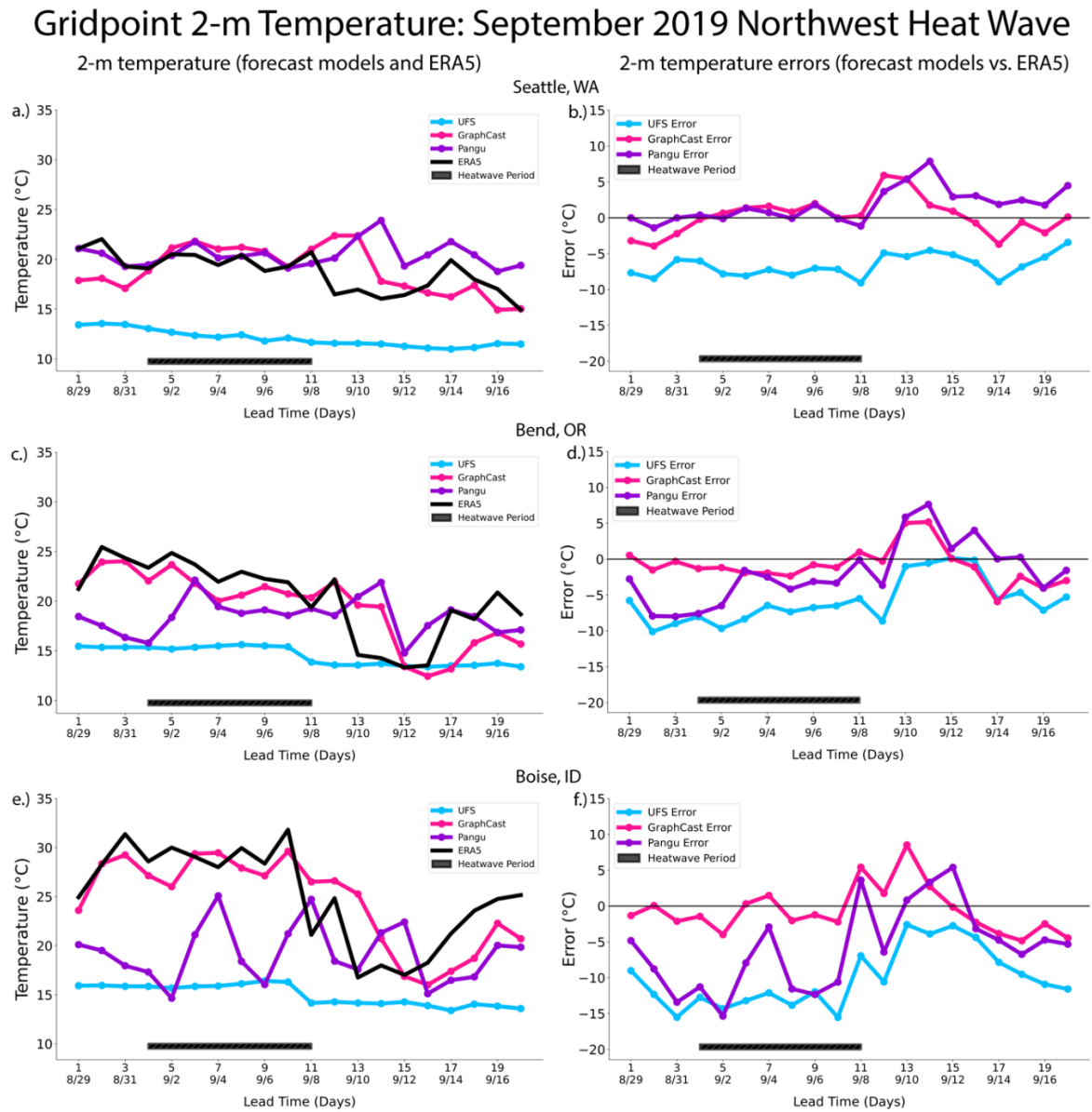

Fig. 6: As in Fig. 5, but for the September 2019 Northwest heat wave.

The GraphCast error behavior for the December 2015 Northeast event mirrors the September 2023 Midwest case, but with larger absolute errors during the heat wave period (5–10 °C vs. 2 – 4°C) and in the overall averages (2.7 °C vs. 1.4 °C) (Figs. 7d, f). Specifically, GraphCast exhibits small errors before the heat wave, has a cold bias during most of the heat wave period, and exhibits a warm bias near the end of the heat wave (Fig. 7f). There is similar error behavior in Pangu during and after the heat wave period, but Pangu has a large warm bias just prior to heat wave onset (Fig. 7f). Pangu for the Northeast event is the only case in which either model has a large warm bias prior to the heat wave, perhaps reflective of it initiating event onset too soon.

GraphCast and Pangu errors for the May 2004 Southeast heat wave (Figs. 7g,h) agree well with the August 2011 event in the same region (Fig. 4). Both models exhibit a cold bias throughout the entire forecast period, with errors larger after heat wave onset than before it (Fig. 7h). As in the August 2011 event, GraphCast consistently outperforms Pangu, especially near heat wave onset. The average GraphCast absolute error (1.9 °C) is the second lowest of the four events (after September 2023 Midwest), and like the cases in Fig. 3, GraphCast performs better in the Southeast than the Northwest.

We next examine the performance of the AIWP models on a seasonal average basis, with an approximately equal number of cases in all four regions. Fifteen heat waves (see section 2c) per season are chosen for analysis. In each season, eight heat waves are during the model training years (1979–2017) and seven heat waves during the model testing years (2018–2023).

For boreal summer, GraphCast and Pangu both exhibit consistent cold biases throughout the forecast period, regardless of whether the event is during the training or testing years (Figs. 8a,b). The largest errors (cold biases) occur near heat wave onset, consistent with our case study results. Overall, GraphCast substantially outperforms Pangu during boreal summer, especially leading up to and during the heat wave period. Boreal spring results (Figs. 8g,h) mirror our summer results (Figs. 8a,b), in that GraphCast substantially outperforms Pangu and there is a consistent cold bias throughout the forecast period in both models, regardless of whether it is the training or testing years.

In boreal winter, Pangu exhibits a warm bias leading up to heat wave onset, which is substantially larger during the training period (Figs. 8e,f). GraphCast and Pangu absolute errors are similar for most of the forecast period, although the GraphCast errors are manifested through a cold bias, the opposite of Pangu. Finally, boreal autumn results (Figs. 8c, d) exhibit an early cool bias similar to our summer results (Fig. 8b), but a post-heat wave warm bias more like our winter results (Fig. 8f).

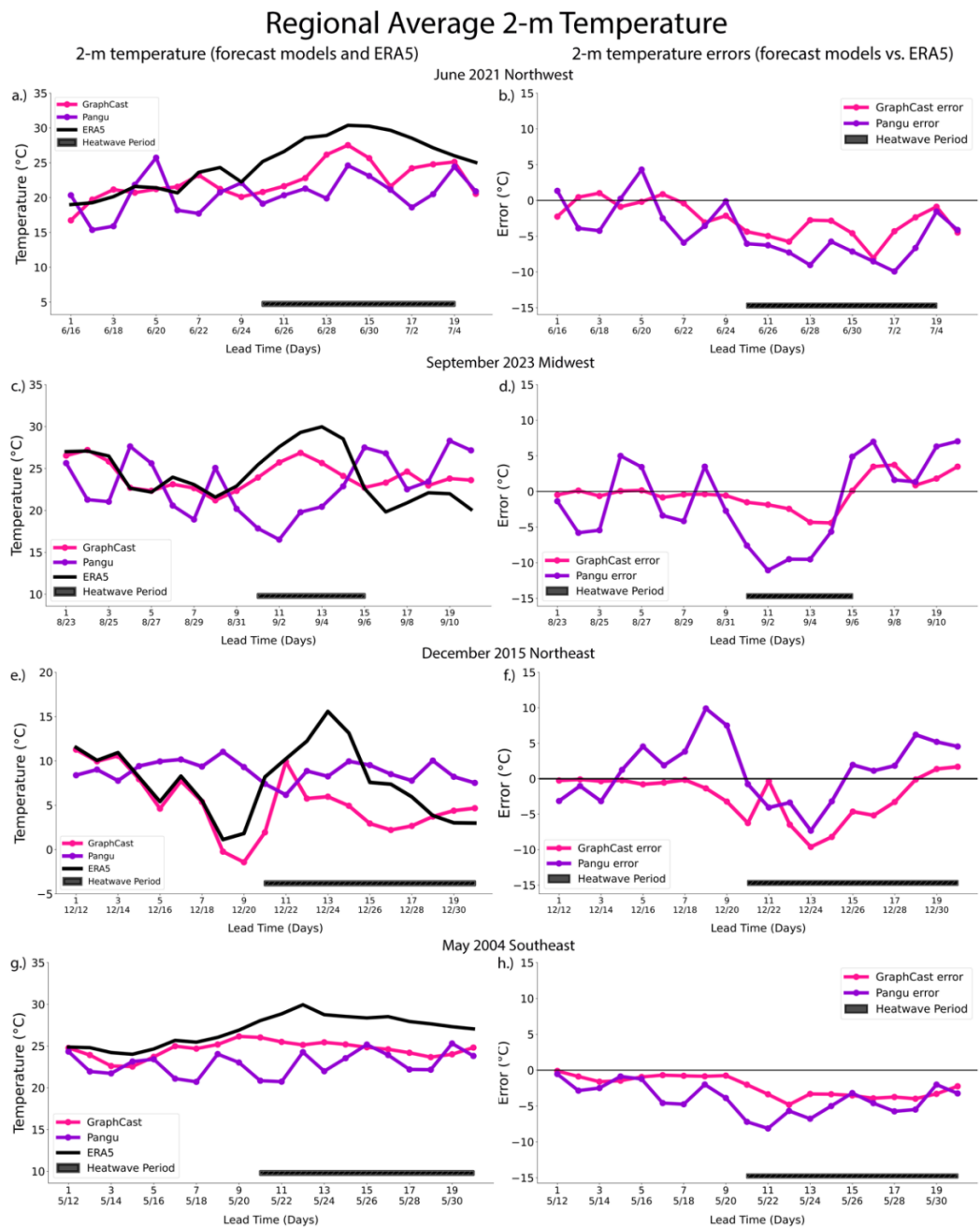

Fig. 7: Regional average (left) 2-m temperature (°C) and (right) 2-m temperature errors (°C) in GraphCast (magenta line) and Pangu (purple line) for the (a,b) June 2011 Northwest heat wave; (c,d) September 2023 Midwest heat wave; (e,f) December 2015 Northeast heat wave; and (g,h) May 2004 Southeast heat wave. Temperature errors are calculated using ERA5 (black line on left-hand panels) as truth and are shown as a function of time. The heat wave period is illustrated by the black line at the bottom of each panel.

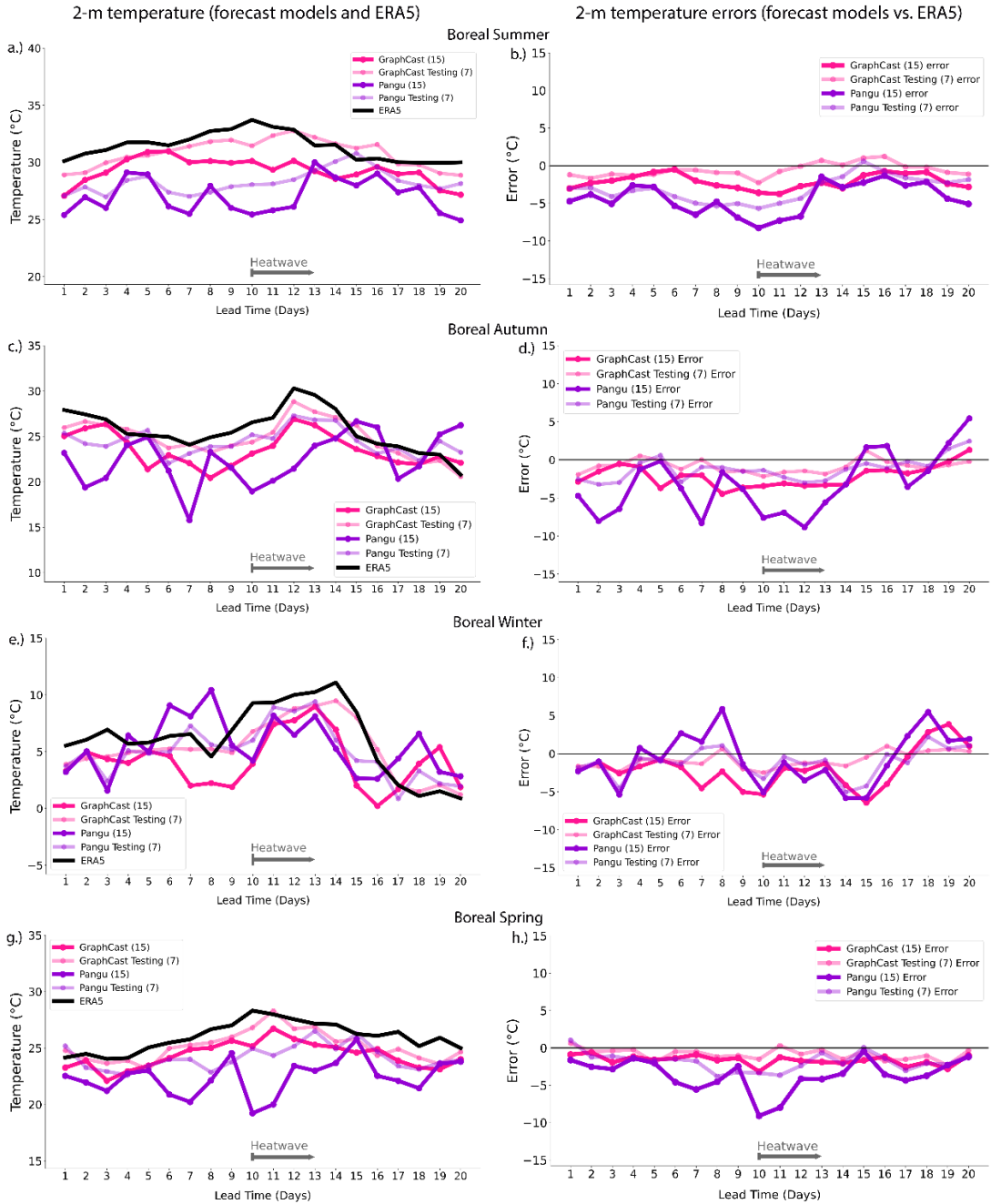

Fig. 8: Seasonal average (left) 2-m temperature (°C) and (right) 2-m temperature errors (°C) for heat waves during the GraphCast training (1979–2017; magenta line) and testing (2018–2023; light pink line) periods, and Pangu training (1979–2017; purple line) and testing (2018–2023; lavender line) periods. Temperatures and temperature errors are averaged over eight training and seven testing period heat waves during (a,b) boreal summer; (c,d) boreal autumn; (e,f) boreal winter; and (g,h) boreal spring. Temperature errors are calculated using ERA5 (black line on left-hand panels) as truth and are shown as a function of time. The heat wave period starts at lead time 10, as illustrated by the black line at the bottom of each panel.

To provide spatial context to the regional average results shown throughout this section, Fig. 9 shows GraphCast (Fig. 9a) and Pangu (Fig. 9b) 2-m temperature errors at lead time 10 for all heat waves across the four selected NCA regions. To create Fig. 9, in each region we first average lead time 10 forecasts over all events; lead time 10 is the first day of the heat wave in each event. Next, we plot the average lead time 10 temperature error for each grid point within a respective region (Fig. 9). For example, the 2-m temperature errors shown in the Northwest are lead time 10 errors at each grid point averaged over the 16 total heat waves (i.e., heat waves that fall within both the model testing and training years) in that region. It is evident that both GraphCast and Pangu have the tendency to underestimate 2-m temperatures during heat waves, particularly within the Northwest region (Figs. 9a,b). In contrast, the AIWP models predict slightly warmer temperatures compared to ERA5 within the mountainous areas of the Carolinas in the Southeast region. This discrepancy may arise from the relatively coarse representation of terrain features in the AIWP models in this study. Furthermore, our analysis demonstrates that Pangu's temperature errors (Fig. 9b) generally exceed those of GraphCast. GraphCast exhibits a specific tendency to overpredict temperatures along the Gulf Coast, particularly in Louisiana and Alabama (Fig. 9a), but still generally outperforms Pangu.

## 4. Discussion and conclusions

This study evaluates the 2-m temperature forecast performance of UFS GEFS and two AIWP models (GraphCast and Pangu) for a set of heat waves across four NCA regions. For our two case studies (Figs. 3–6), GraphCast and UFS GEFS exhibit similar performance in our regional average results (Figs. 3,4). The UFS GEFS results throughout our case studies are mixed: There are small warm biases in the regional averages, especially in the Northwest (Fig. 3), but large cold biases at specific locations throughout the Northwest and Southeast (Figs. 5,6). To provide additional context, in a study of UFS boreal summer forecast skill, Seo et al. (2024) report that UFS has a consistent warm bias in the western CONUS at all lead times. However, for annual mean 2-m temperature, Stefanova et al. (2022) show a large mean cold bias over the entire CONUS at all lead times. In an examination of boreal summer extreme events, Krisnamurthy and Stan (2022) show large UFS errors for 2-m temperature in both the Southeast and Northwest with large sub-regional variability throughout the western CONUS, especially at longer lead times. Previous studies suggest several potential reasons for large UFS 2-m temperature forecast errors, especially for extreme heat events at longer lead times and over the Western CONUS: biases in the slowly varying modes of climate variability in the Pacific Ocean (e.g., ENSO, Krisnamurthy and Stan 2022; Stan et al. 2023), tropical Pacific convection and the Madden-Julian Oscillation (Choi and Stan 2025), as well as land-atmosphere coupling and soil moisture (Benson and Dirmeyer 2023; Seo et al. 2024). Although it is beyond the scope of this study to investigate the causes of the UFS GEFS errors in our case studies, future work could examine each of these potential mechanisms as a source of forecast error during heat waves.

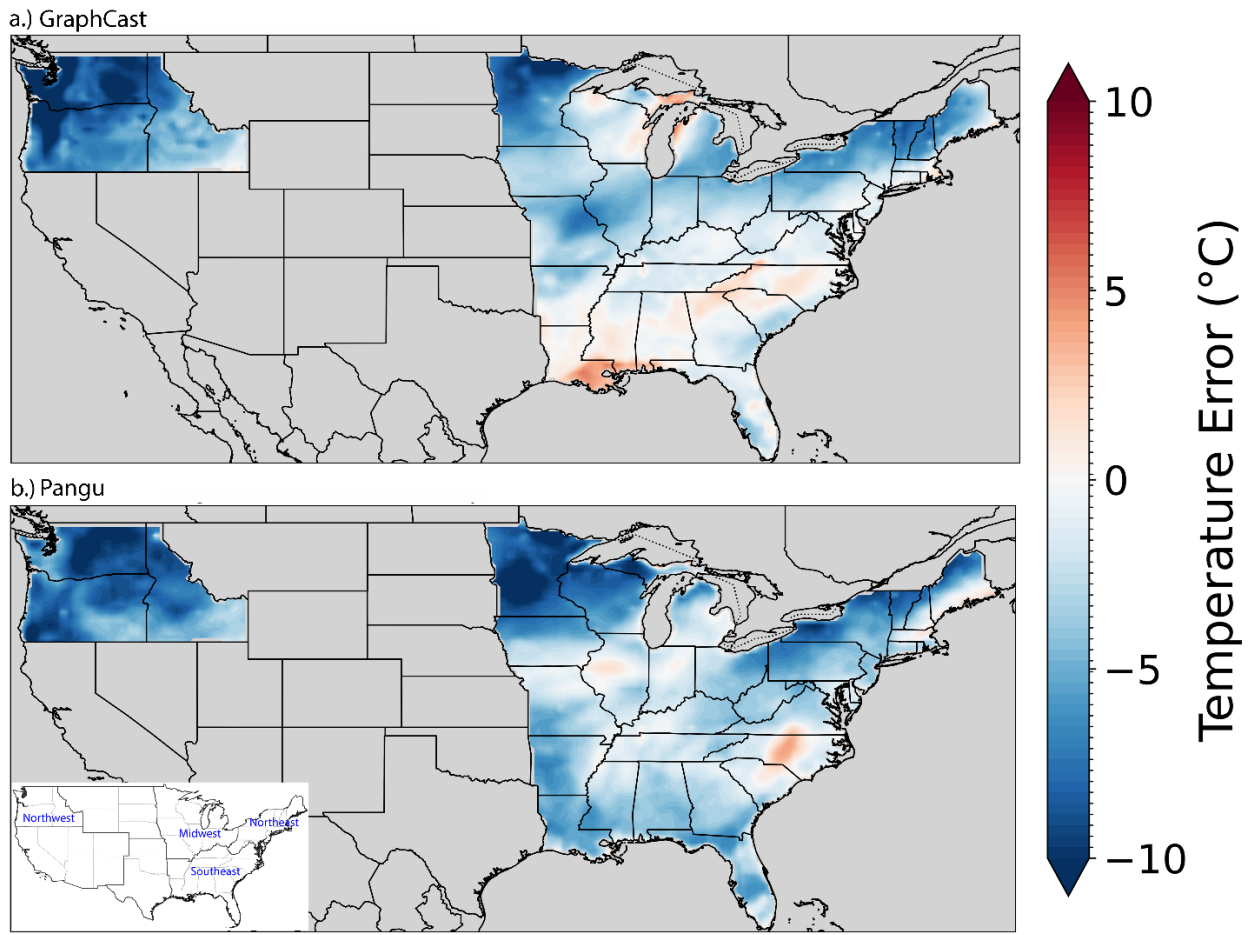

Fig. 9: For the four NCA regions (Northwest, Midwest, Northeast, Southeast), spatial distribution of 2-m temperature errors (°C, shaded) averaged over all heat wave events across all boreal seasons for (a) GraphCast and (b) Pangu. The lower left inset shows a map of the seven NCA regions (black outlines; USGCRP 2023) with the four regions used in this study labeled in blue, also shown in Fig. 1.

Overall, GraphCast performs quite well in all seasons, albeit with a persistent small cold bias prior to and during the heat wave period (Fig. 8). Pangu does better in the seasonal evaluation (Fig. 8) than the case studies (Figs. 4–7). In all four seasons, both models consistently exhibit smaller errors during the testing period compared to the training years (Fig. 8). However, we emphasize that the total number of heat waves in our study is relatively small, and the comparison between the testing and training years would need to be expanded to make broad conclusions. Finally, both GraphCast and Pangu exhibit consistent cold biases throughout all seasons except boreal winter, especially leading up to heat wave onset (Fig. 8). To that point, boreal winter has two unique findings among the four seasons: 1) Pangu has a warm bias prior to the heat wave, suggesting too early an onset, and 2) both AIWP models exhibit warm biases after the heat wave ends (Fig. 8), suggesting that they do not decay the heat wave quickly enough. Future work should investigate the potential causes of warm biases being predominantly unique to winter.

As mentioned in section 2, the Pangu developers (Bi et al. 2023) introduce hierarchical temporal aggregation for medium-range forecasts and suggest using the 24-h model for longer lead times. The hierarchical temporal aggregation is intended to reduce the number of iterations required to train a series of models with longer lead times, thereby potentially reducing medium range forecast errors (Bi et al. 2023). There are four initialization times for the 24-h model outputs (00, 06, 12, and 18 UTC) and each forecast is instantaneous, not a daily average. In other words, one must choose an initialization time and use the 24-h model from that time. Therefore, in this study we use the Pangu 6-h model, because UFS GEFS and GraphCast both output on a 6-h basis and an instantaneous value for a specific forecast hour would not be consistent with the uniform daily averages of UFS GEFS and GraphCast. Furthermore, Pasche et al. (2025) warn that most AIWP models use a large autoregressive time step (i.e., coarse temporal resolution) that can impact forecasts in which the daily maximum and/or minimum values of a forecast parameter are important, such as 2-m temperature during heat waves. Pasche et al. (2025) stipulate that even a 6-h time step can miss such peaks and valleys, lending even further credence to our choice not to focus on the Pangu 24-h model. Alternatively, we could average Pangu 24-h output at each of the four daily initialization times and create daily averages from those. However, doing so would not preserve consistency with how we create the daily averages (using 6-h output) of UFS GEFS and GraphCast, particularly for the 20-day forecast period we evaluate. Our results using the 6-h Pangu output (Figs. 3–8) show that Pangu errors largely exceed GraphCast errors across most cases, regions, and seasons. In addition, Pangu exhibits a "see-saw" effect, where the errors have large oscillations within a given 24-h period (see example in Fig. 3b compared to UFS GEFS and GraphCast. These oscillations are likely directly related to our choice to use the 6-h output. For more insight into the differences between the 6-h and 24-h Pangu output, Fig. S2 shows results for the September 2019 Northwest heat wave.

In summary, extreme heat is the deadliest weather-related hazard in the United States, and is becoming more frequent, intense, and longer duration over time. While heat wave predictability has improved, forecast skill still lags behind for medium-range and S2S timescales, necessitating technological advances and novel approaches. This study investigates the abilities of two AIWP models (GraphCast and Pangu) to forecast 2-m temperature extremes and associated heat waves across four CONUS regions for medium-range and subseasonal timescales. We also compare AIWP model temperature forecast skill to that of one traditional physics-based S2S NWP model (UFS GEFS). As AIWP models are relatively new and are not always able to accurately represent physical processes, investigating their forecast skill along with a comparison to a traditional NWP model is vital to understanding whether AIWP models can offer advantages in the predictability of extremes (e.g., heat waves). Our results demonstrate that GraphCast and Pangu show promise and skill advances relative to UFS GEFS by some measures. While more work is required to better understand the limitations and driving mechanisms of AIWP forecast skill, results such as those for GraphCast in this study offer promise that these new tools can result in a major leap forward for the medium-range and S2S predictability of heat waves and other extreme weather events.

*Acknowledgements.*

This research was supported by a NOAA Office of Atmospheric Research (OAR) grant (NA22OAR4310621) and a Heising-Simons Foundation grant (#2023-4720). MAF was funded by the Regional and Global Model Analysis program area of the U.S. Department of Energy's (DOE) Office of Biological and Environmental Research (BER) as part of the Program for Climate Model Diagnosis and Intercomparison Project. The authors thank ECMWF for making ERA5 data publicly available, NOAA for doing the same with UFS GEFS, and the creators of Google GraphCast and Pangu-Weather for making their models accessible through GitHub. We also thank Dr. Shawn Milrad of Embry-Riddle Aeronautical University for providing technical guidance and editing an earlier version of this manuscript.

*Data Availability Statement.*

ERA5 data can be accessed from the Analysis-Ready, Cloud Optimized (ARCO) repository on Google Cloud (https://console.cloud.google.com/marketplace/product/bigquery-public-data/arco-era5). The UFS GEFS data can be downloaded using the Amazon Web Services repository (https://noaa-gefs-retrospective.s3.amazonaws.com/index.html). Finally, Google GraphCast (https://github.com/google-deepmind/graphcast) and Pangu-Weather (https://github.com/198808xc/Pangu-Weather) are available through their respective GitHub sites.

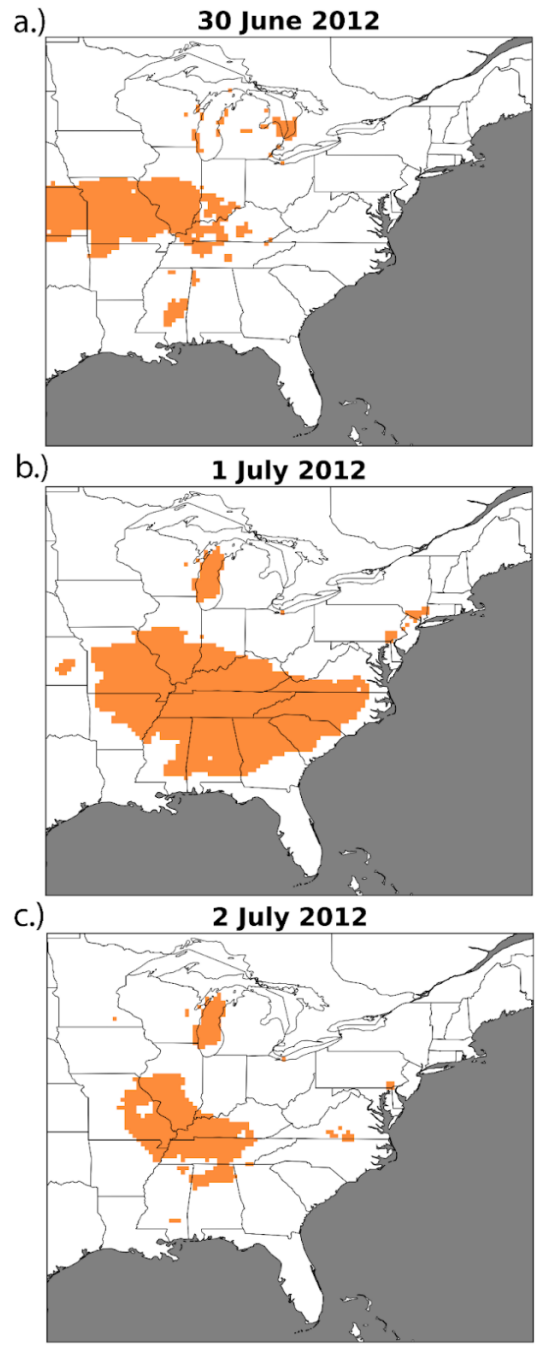

Fig. S1: For the 2012 heat wave that impacted both the Midwest and Southeast NCA regions, the orange shading indicates areas where daily mean 2-m temperatures exceeded the 95[th] percentile on (a) 30 June and (b) 1 July, and (c) 2 July.

# Regional Average 2-m Temperature

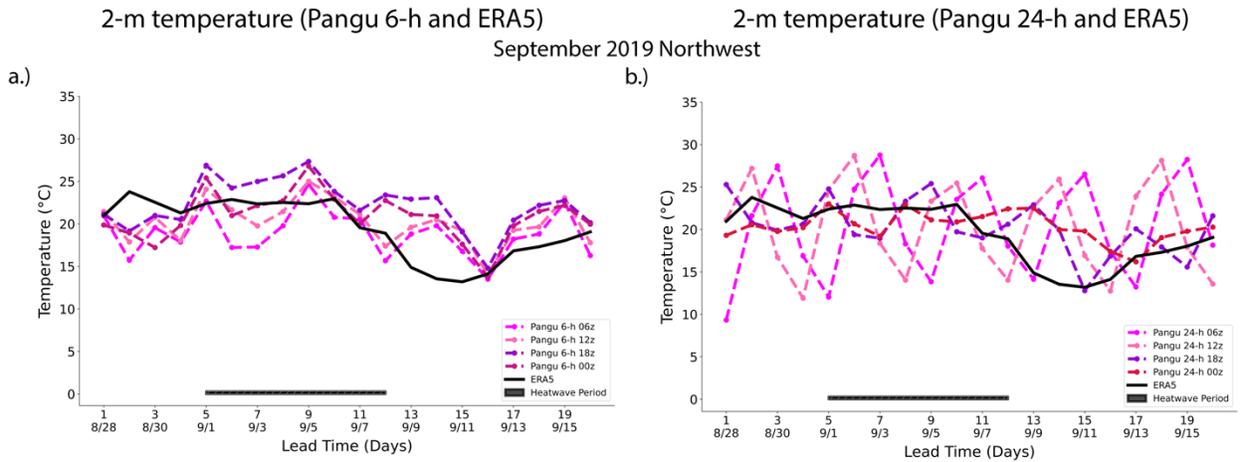

Fig. S2: For the September 2019 Northwest heat wave, regional average 2-m temperature (°C) for (a) four runs (each six hours apart) of the Pangu 6-h model and (b) four runs (each six hours apart) of the Pangu 24-h model. In each panel, the 00 UTC run is plotted in dashed red, the 06 UTC run in dashed magenta, the 12 UTC run in dashed pink, and the 18 UTC run in dashed purple. In both panels, ERA5 2-m temperatures (°C, solid black lines) are plotted as truth, and the heat wave period is illustrated by the black line at the bottom.